\documentclass [prb,twocolumn,superscriptaddress,floatfix,amsmath,amssymb] {revtex4-2}
\usepackage{amsmath}
\usepackage{graphicx}
\usepackage{hyperref}
\usepackage{color}
\usepackage{amssymb}
\usepackage{bm}
\usepackage{wasysym}

\newcommand{\beq} {\begin{equation}}
\newcommand{\eeq} {\end{equation}}
\newcommand{\bea} {\begin{eqnarray}}
\newcommand{\eea} {\end{eqnarray}}
\newcommand{\be} {\begin{equation}}
\newcommand{\ee} {\end{equation}}

\definecolor{darkgreen}{RGB}{0,170,0}

\begin{document}
\title{Non-BCS behavior of the pairing susceptibility near the onset of superconductivity in a quantum-critical metal}

\author{Artem Abanov}
\affiliation{Department of Physics, Texas A\&M University, College Station,  USA}
\author{Shang-Shun Zhang}
\affiliation{Department of Physics and Astronomy, The University of Tennessee, Knoxville, Tennessee 37996, USA}
\author{Andrey V. Chubukov}
\affiliation{School of Physics and Astronomy and William I. Fine Theoretical Physics Institute, University of
Minnesota, Minneapolis, MN 55455, USA}

\date{\today}

\begin{abstract}
We analyze the dynamical pairing susceptibility $\chi_{pp} (\omega_m)$ at $T=0$ in a quantum-critical metal, where superconductivity
  emerges out of a non-Fermi liquid ground state once the pairing interaction exceeds a certain threshold.    We obtain $\chi_{pp} (\omega_m)$ as the ratio of the fully dressed dynamical pairing vertex $\Phi (\omega_m)$ and the bare $\Phi_0 (\omega_m)$ (both infinitesimally small).
    For  superconductivity out of a Fermi liquid,
    the pairing susceptibility is positive above
       $T_c$,
        diverges at
        $T_c$,
         and becomes negative below it. For superconductivity out of a non-Fermi liquid, the behavior of $\chi_{pp} (\omega_m)$ is different in two aspects: (i) it  diverges at the onset of pairing  at $T=0$ only for a certain subclass of bare $\Phi_0 (\omega_m)$ and remains non-singular  for other $\Phi_0 (\omega_m)$, and (ii)
         below the instability,  it becomes a non-unique function of a continuous parameter $\phi$
         for an arbitrary $\Phi_0 (\omega_m)$. The susceptibility is negative in some range of $\phi$ and diverges at the boundary of this range.   We argue  that this behavior of the susceptibility  reflects a multi-critical nature of a
          superconducting transition in a quantum-critical metal when immediately below the instability an infinite number of superconducting states emerges
   simultaneously with different amplitudes of the order parameter down to an infinitesimally small one.
     %
       %
   \end{abstract}
\maketitle

\section{Introduction}
\label{sec:1}

Pairing of incoherent fermions out of a non-Fermi liquid ground state is a fascinating subject that  attracted  significant attention over the last three decades.  Examples of such pairing include fermions at half-filled Landau levels~\cite{PALee1989,*Monien1993,*Nayak1994,*Altshuler1994,*Kim_1994,Fradkin2016}, quantum-critical
 metals at the onset of  various spin and charge orders~\cite{Millis1992,Altshuler1995a,Sachdev1995,acf,acs,*acs2,*finger_2001,*acn,Subir,*Subir2,*Sachdev2019,vojta,
 *Sachdev_22,
wang,efetov,*efetov2,*efetov3,*benlagra2011luttinger,metlitski2010quantum1,*metlitski2010quantum2,*max_2,*max_last,
sslee2,*sslee_2018,Bauer2015,ital,*ital2,*ital3,wang,*wang23,wang_2,*wang_22,berg,*berg_2,*berg_3,*Hu2017,*berg_4,*meng,matsuda,*peter,
khodas,*tsvelik,*tsvelik_1,*tsvelik_2,*tsvelik_3,Metzner2003,*DellAnna2006,metzner,*metzner_1,*metzner_2,tremblay_2,She2009,
*Yang2011,Wolfle2014,Varma_2016,georges,Kotliar2018,*Wu_19,avi,*avi1,*avi2,*avi_5,Kumar_24,nosov,*foster},
 electron-phonon systems at small Debye frequency~\cite{Scalapino_69,*scal,Marsiglio_91,*ad,*combescot,Mirabi_2020,*boyack,esterlis,*Chubukov_2020b,prokofiev,secchi2020phonon,emil,*emil1,*emil2,
 *emil3,*emil4,math,*math_3,ss_fe,*berg_a},
SYK-type systems~\cite{Patel2019,Schmalian_19,Wang_19,schmalian_19a,Chowdhury_2020,Classen_a}, e.g.,  quantum dots,  and systems near a van Hove singularity, particularly a higher-order one~\cite{Classen_Betouras,*Chamon,*Risto}.  Rich physics here comes from the fact that non-Fermi liquid and superconductivity are competing phenomena. Namely, a fermionic incoherence  in a non-Fermi liquid  weakens the system's ability to create fermionic pairs and  develop a coherent supercurrent.  At the same time,  if superconductivity develops, the associated gap opening reduces the scattering at low energies and renders fermionic coherence.
   The outcome of the competition
   depends on the interplay between the
  interaction in the particle-hole and particle-particle channels as the first
   gives rise for a non-Fermi liquid while the second
    gives rise to pairing. In most cases, the two interactions
   originate from the
     same effective 4-fermion interaction between low-energy fermions
      and are comparable in strength.
  In this situation, non-Fermi liquid and superconductivity compete on equal footing.
   In the quantum-critical models studied so far~\cite{raghu_15,*raghu2,*raghu3,*Fitzpatrick_15,moon_2,*Subir2,WAAYC,paper_1,*paper_2,*paper_3,*paper_4,*paper_5,*paper_6,*odd,*jetp,math_2}  superconductivity
    wins, but  by purely numerical reasons.

 In this communication we consider a more generic case, when the interactions in the particle-hole and particle-particle channel  have the same functional form, but
  differ by some factor.
 Such a situation emerges~\cite{raghu_15}
  upon extension of the original $SU(2)$ model to matrix $SU(N)$ with $N \gg 1$:
  the interaction in the
  particle-particle channel becomes smaller
  by the factor of $N$.  A similar situation holds in Yukawa-SYK models with the only difference that the relative smallness of the particle-particle interaction is controlled by a continuous parameter -- the strength of time-reversal symmetry breaking disorder~\cite{Schmalian_19,Wang_19,schmalian_19a,Classen_a}. In our study we  follow these earlier works. We label the ratio of the two interactions by $1/N$, in analogy with the $SU(N)$ case, but
   treat $N$ as a continuous parameter, like in Yukawa-SYK models.

\begin{figure}[!t]
\centering
\includegraphics[angle=0,width=0.5\textwidth]{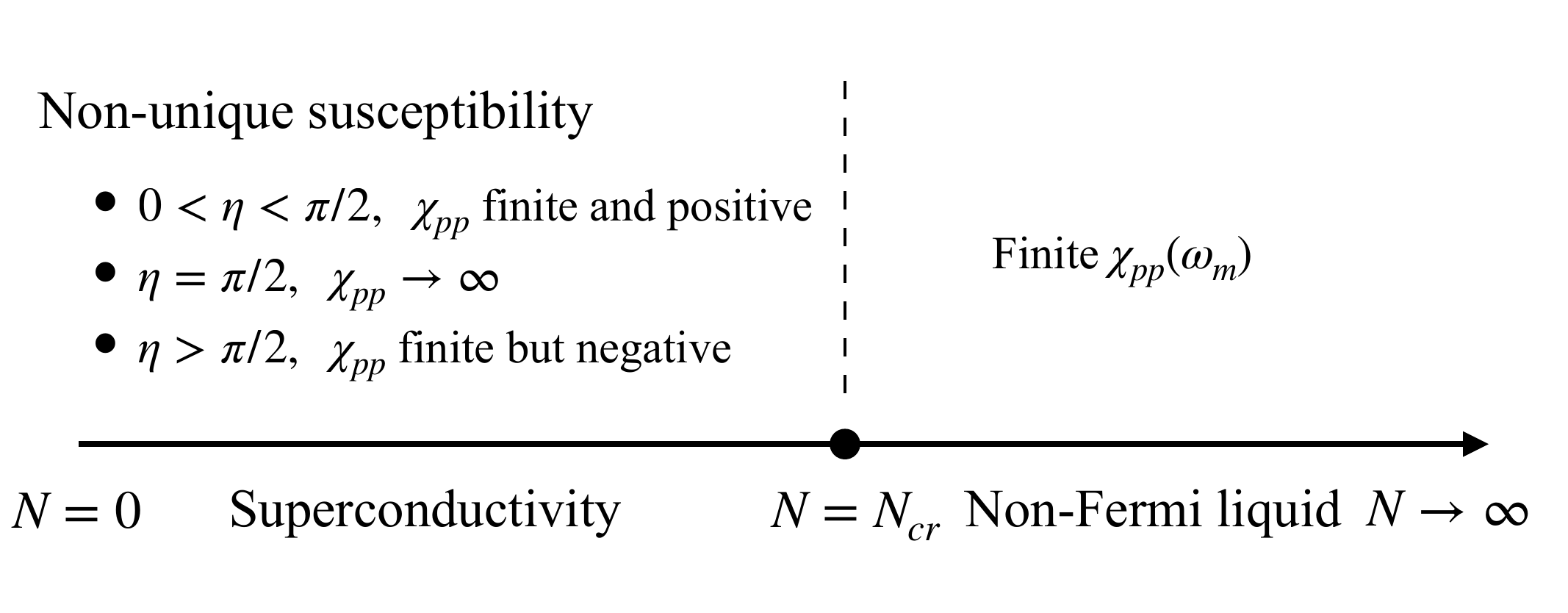}
\caption{A schematic zero-temperature phase diagram of the $\gamma$-model, extended to $N >1$.  A non-Fermi liquid phase with a positive and non-singular pairing susceptibility holds for $N > N_{\text{cr}}$, and a superconducting phase holds for $N < N_{\text{cr}}$. In the latter, the pairing susceptibility is non-unique. The critical value $N_{\text{cr}}$ varies with $\gamma$ between $N_{cr} =\infty$ at $\gamma =0$ and $N_{cr} =1$ for $\gamma =1$. }
\label{fig:fig1}
\end{figure}

To simplify the presentation, we focus on superconductivity in a  metal near a quantum-critical point.
  Earlier studies\cite{raghu_15,*raghu2,*raghu3,*Fitzpatrick_15,khvesh,WAAYC,paper_1,*paper_2,*paper_3,*paper_4,*paper_5,*paper_6,
  *odd,*jetp,Chubukov_2020a,
  Wang_H_17,*Wang_H_18,math_2} of quantum-critical metals  have found that to a good accuracy (analytical in some cases, numerical in others) the  competition between non-Fermi liquid and superconductivity in a spatial channel with the highest attraction  can be can described by an effective Eliashberg-type theory  with local dynamical interaction  $V(\Omega_m)$, local self-energy $\Sigma (\omega_m)$ and local pairing vertex $\Phi (\omega_m)$ (all three are real functions of frequency on the Matsubara axis). A more familiar superconducting gap function $\Delta (\omega_m)$ is related to $\Phi (\omega_m)$ as
   $\Delta (\omega_m) = \Phi (\omega_m)/(1 + \Sigma (\omega_m)/\omega_m)$. These earlier studies have shown
 that at $T=0$ there exists a critical $N_{cr}$ separating regions of a non-Fermi liquid ground state at $N > N_{cr}$ and a superconducting state at $N < N_{cr}$ (see Fig.~\ref{fig:fig1}).
    A conventional wisdom would then tell  that the
     dynamical  pairing vertex $\Phi (\omega_m)$ is zero
     at $N > N_{cr}$, is infinitesimally small at
      at $N =N_{cr}-0$, and is finite at
       $N < N_{cr}$ with magnitude increasing
        with deviation from $N_{cr}$.   The actual situation is, however, more complex as it turns out that $N = N_{cr}$ is a multi-critical point, below which an infinite number of topologically distinct solutions $\Phi_n (\omega)$ appears simultaneously with $n =0,1,2...$ (Refs. \cite{paper_1,son_1,*son_2,*son_3,Classen_a}   The function $\Phi_n (\omega_m)$ has $n$ zeros along the Matsubara axis in the upper half-plane of frequency.
        \footnote{This gives rise to $n$ phase slips of the phase of the complex
         gap function on the real axis: $\psi_n (\infty) - \psi_n (0)) = n\pi$, where
        $\Phi_n (\omega) = |\Phi_n (\omega)| e^{i\psi_n (\omega)}$.  The variation of $\psi_n (\omega)$  can be extracted from photoemission and other measurements.}
         The magnitude of $\Phi_n (\omega_m)$ decreases with $n$ and becomes infinitesimally small at $n \to \infty$ for any $N < N_{cr}$.  This implies that the
         linearized equation for $\Phi(\omega_m)$ has a solution not only at $N = N_{cr}-0$ but also for arbitrary  $N < N_{cr}$, along with an infinite set of solutions of the non-linear gap equation.

    This highly unconventional behavior calls for the analysis of the pairing susceptibility, $\chi_{pp} (\omega_m)$.
     We define $\chi_{pp}$ as the system reaction to an  infinitesimally small input $\Phi_0 (\omega_m)$, i.e., as $\chi_{pp} (\omega_m) = \Phi (\omega_m)/\Phi_0 (\omega_m)$, where $\Phi (\omega_m)$ is the fully dressed response function -- the solution of the gap equation induced by $\Phi_0 (\omega_m)$. In a BCS superconductor, $\chi_{pp} (\omega_m) = \chi_{pp}$  is frequency independent. It is  positive in the normal state, diverges at the critical point, and becomes negative below the critical point, indicating that the normal state is unstable towards pairing.  In the quantum-critical case, the susceptibility has to reflect the fact that  at $\Phi_0 =0$ and $N < N_{cr}$, there exist a set of solutions with a finite $\Phi_n (\omega_m)$, for which one could expect a negative but finite  pairing susceptibility, and the solution with infinitesimally small $\Phi_\infty (\omega_m)$, for which
       one would  expect the susceptibility to diverge.  The goal of our work is to reconcile these two seemingly different forms of the susceptibility.

     We will also explain another  unconventional feature of the pairing susceptibility, detected in earlier studies by comparing  the quantum-critical pairing with the BCS case~\cite{acf,paper_1,jetp}.
     Namely, in BCS theory,  the pairing susceptibility is obtained by summing up ladder series of Cooper logarithms at a finite temperature $T$. The series are geometrical and sum up into
      $\chi_{pp} \propto 1/\log{T/T_c}$ implying that it
       diverges at $T= T_c$.  For the quantum-critical case, the ladder series for $\chi_{pp}$ at $T=0$ are also logarithmic and contain powers of $\log{|\omega_m|}$ due to a combination of singular interaction and singular self-energy (we will present explicit expressions below).
        However, the ladder series are not geometrical and sum up into a power-law form
        $ \chi_{pp} (\omega_m) \propto \left(\frac{1}{|\omega_m|}\right)^{\frac{1-\gamma}{N}}$.
       There is no indication in this formula that a superconducting instability develops at $N = N_{cr}$.   We argue that
        an instability can be detected, but one has to analyze the validity of perturbation theory.

In this work, we  first consider the case when the bare $\Phi_0 (\omega_m) = \Phi_0$ is independent on frequency.
 We show that
 the behavior of the pairing susceptibility is highly unconventional: it does not diverge as $N$ approaches $N_{cr}$ from above and remains finite and positive  at $N = N_{cr}$.
 However, once $N$ gets smaller than $N_{cr}$
  the perturbative expansion in  $\log{|\omega_m|}$  breaks down.
  To see this, we will go beyond the leading logarithms and collect non-logarithmic corrections in powers of $1/N$  for each term in the logarithmic series.
  We show that at $N > N_{cr}$, the pairing susceptibility becomes the function not only of frequency but also of a running parameter $\eta$.  For a particular value of $\eta=\pi/2$, $\chi_{pp} (\omega_m, \pi/2)$ diverges for all $\omega_m$,  consistent with the existence of an infinitesimally small $\Phi_\infty (\omega_m)$. For the range of $\eta$ near $\pi/2$, the pairing susceptibility is finite, but negative,  consistent with the existence of the set of finite $\Phi_n (\omega_m)$.

 We next  analyze in detail a generic case when the input infinitesimal $\Phi_0$ is a function of frequency, $\Phi_0 (\omega_m)$.   We show that the behavior is generally the same as for a constant $\Phi_0$, i.e., there is an abrupt change of $\chi_{pp}$ at $N = N_{cr}$ from a regular  function of frequency to a  function of two variables - frequency and  a running parameter.  However for a set of $\Phi_0 (\omega_m)$, $\chi_{pp}$ diverges as $N$ approaches $N_{cr}$  from above, and at $N < N_{cr}$ again becomes a function of a running parameter.

 The outline of the paper is the following. In the next section we present the model and the set of coupled equations for the pairing vertex and the self-energy.  We briefly describe the results obtained by solving these equations without the source term  and then present the equation for the pairing susceptibility.
   In Sec. \ref{sec:3} we analyze the pairing susceptibility for the case when the source term $\Phi (\omega_m) = \Phi_0$ in independent on frequency. We show that the pairing susceptibility remains finite for $N \geq N_{cr}$, and becomes a multi-valued  function at $N < N_{cr}$. In Sec. \ref{sec:4} we extend the analysis to frequency dependent $\Phi_0 (\omega_m)$. We show that for a generic $\Phi_0 (\omega_m)$ the behavior of the susceptibility is similar to the case of a constant $\Phi_0$, but for a special $\Phi_0 (\omega_m)$,  the pairing susceptibility diverges at $N \to N_{cr}$ from above, and again becomes a multi-valued function at $N <N_{cr}$.
   In Sec. \ref{sec:5} we present the explanation of the  behavior of the pairing susceptibility at $N \geq N_{cr}$
    by invoking the criterium of normalizability of $\Phi (\omega_m)$ and show that in physical terms it implies that the condensation energy must remain finite.   We present our conclusions in Sec. \ref{sec:6}.

 \section{Model and gap equation}
\label{sec:2}

 In a quantum critical metal the dynamical interaction between fermions $V (\Omega_m) = (g/|\Omega_m|)^\gamma$ is  mediated by massless fluctuations of a collective bosonic degree of freedom in either spin or charge channel (the   $\gamma-$model, named by he exponent).
  As we said, we assume that the interactions in particle-hole and  particle-particle channels have the same functional form, but the latter an extra factor $1/N$.

 Taken alone, the interaction in the particle-particle channel gives rise to a superconducting ground state  with a non-zero $\Phi (\omega_m)$, whose spatial symmetry is specified by the underlying microscopic model (e.g., $d-$wave for the quantum-critical metal at the onset of an antiferromagnetic order). In turn, the interaction in the particle-hole channel, taken alone, gives rise to a non-Fermi liquid ground state with the self-energy
 \bea
 \Sigma (\omega_m) &=& \text {sign}(\omega_m) \Sigma^* (\omega_m), \nonumber \\
 \Sigma^* (\omega_m) &=&  \int_0^{|\omega_m|} V(\Omega_m) d\Omega_m = |\omega_{m}|^{1-\gamma} \omega^\gamma_0
 \label{eq:5}
\eea
where $\omega_0 = g/(1-\gamma)^{1/\gamma}$.
 Below we will measure the self-energy, the pairing vertex, and frequency  in units of $\omega_0$, i.e., redefine  $\Sigma (\omega_m) \to \omega_0 \Sigma (\omega_m)$,   $\Phi (\omega_m) \to \omega_0 \Phi (\omega_m)$, and   $\omega_m \to \omega_m/\omega_0$.

 The ground state of the $\gamma$ model in the presence of both interactions has been analyzed before.
 To address the interplay between superconductivity and non-Fermi liquid, one has to solve the set of two coupled non-linear equations for the pairing vertex and the self-energy:
 \bea
&&\Phi (\omega_m) = \frac{1-\gamma}{2N}\int \frac{d\omega^{'}_m \Phi (\omega^{'}_m)}{
\sqrt{\left(\omega^{'}_m + \Sigma (\omega^{'}_m)\right)^2 + \Phi^2 (\omega^{'}_m)}} \frac{1}{|\omega_m -\omega^{'}_m|^\gamma} \nonumber \\
&&\Sigma(\omega_m) =
\frac{1-\gamma}{2}
\int \frac{d\omega^{'}_m \Sigma (\omega^{'}_m)}{
\sqrt{\left(\omega^{'}_m + \Sigma (\omega^{'}_m)\right)^2 + \Phi^2 (\omega^{'}_m)}} \frac{1}{|\omega_m -\omega^{'}_m|^\gamma} \nonumber \\
&& \label{eq:7a}
\eea
Similar, though not identical equations have been obtained for the Yukawa SYK model of dispersion-less fermions, interacting with Einstein phonons, in the limit when the number of fermionic and bosonic flavors tend to infinity, but their ratio is finite.

 Eqs. (\ref{eq:7a}) have been analyzed before\cite{paper_1}, both analytically and numerically. We do not discuss the details  and just list the results, which will serve as an input for the analysis of the susceptibility.
These results are different for $0<\gamma <1$ and $\gamma >1$ (Ref. \cite{paper_4}
 For definiteness,  we focus on  $0<\gamma <1$.

 \begin{itemize}
 \item
 At large $N > N_{cr}$, the ground state remains a non-Fermi liquid, i.e., $\Phi (\omega_m) =0$, and the self-energy is given by (\ref{eq:5}).
 \item
  At $N < N_{cr}$, the ground state is a superconductor with $\Phi (\omega_m) \neq 0$. The feedback from superconductivity renders the Fermi liquid form of the self-energy.
 \item
   The value of  $N_{cr}$ depends on the exponents $\gamma$ as
   \beq
   N_{cr} = \left(\frac{1-\gamma}{2}\right) \frac{\Gamma^2\left(\frac{\gamma}{2}\right)}{\Gamma\left(\gamma\right)} \left(1+ \frac{1}{\cos{\left(\frac{\pi \gamma}{2}\right)}}\right)
 \label{eq:10}
 \eeq
  At small $\gamma$, $N_{cr} \approx 4(1-\gamma)/\gamma$; at $\gamma \to 1$, $N_{cr} \to 1$.  That  $N_{cr}$ is finite already implies that the pairing at a quantum-critical point  is qualitatively different from that in a Fermi liquid, where superconductivity develops already for arbitrary weak attraction because of Cooper logarithm.
   At small $\gamma$, when deviations from the Fermi liquid behavior become relevant only at the smallest frequencies,   $N_{cr} \approx 4/\gamma \gg 1$, i.e., the threshold is still finite, but superconductivity develops when the interaction in the particle-particle channel is still weak.
 \item
 There is an infinite discrete set of solutions of the non-linear gap equation. The solutions, $\Phi_n (\omega_m)$, are labeled by integer $n$. They all emerge at $N =N_{cr}-0$. At small $N_{cr} -N$, $\Phi_n (0) \propto \exp\left(\pi n /(N_{cr} -N)^{1/2} \right)$, resembling the behavior near a BKT transition.  The solutions are topologically different: $\Phi_n (\omega_m)$ has n nodes on the Matsubara axis, and each such nodal point is a center of a dynamical vortex.  On a real axis, the corresponding $\Phi_n (\omega)$ is a complex function,  $\Phi_n (\omega) = |\Phi_n (\omega)|e^{i\eta (\omega)}$ with $n$ phase slips, leading to $\eta (\infty) -\eta(0) = \pi n$.
 \item
  The $n=0$ solution is the true minimum of the ground state energy, all other solutions are saddle points with $n$ unstable directions.   The limiting case $n=\infty$ corresponds to infinitesimally small $\Phi_\infty (\omega_m)$, which is  the solution of the linearized gap equation with $\Sigma (\omega_m)$ given by (\ref{eq:5}). Such solution
  has been explicitly found analytically~\cite{paper_1}
 \end{itemize}

\subsection{Pairing susceptibility}

    Our goal is to obtain the dynamical pairing  susceptibility at zero temperature. For this, we depart from a non-Fermi liquid ground state with $\Sigma (\omega_m)$ given by \eqref{eq:5}, introduce an infinitesimally small pairing vertex $\Phi_0 (\omega_m)$ and solve the gap equation for infinitesimally small $\Phi (\omega_m)$ with  $\Phi_0 (\omega_m)$  acting as a source:
 \beq
\Phi (\omega_m) = \frac{1-\gamma}{2N}\int \frac{d\omega^{'}_m \Phi (\omega^{'}_m)}{|\omega^{'}_m|^{1-\gamma} |\omega_m -\omega^{'}_m|^\gamma} \frac{1}{1+ |\omega^{'}_m|^\gamma} + \Phi_0 (\omega_m)
 \label{eq:7}
\eeq
   The pairing susceptibility $\chi_{pp} (\omega_m)$ is the ratio of $\Phi$ and $\Phi_0$:
\beq
 \chi_{pp} (\omega_m) =  \frac{\Phi (\omega_m)}{\Phi_0 (\omega_m)}.
 \label{eq:6}
\eeq

  For small $\gamma$, when frequency variation in the integrand in the r.h.s of (\ref{eq:7}) is slow, the term $|\omega_m - \omega^{'}_m|^\gamma$ in (\ref{eq:6}) can be approximated by   $|\omega_m|^\gamma$ for $|\omega_m| >|\omega^{'}_m|$ and by $|\omega^{'}_m|^\gamma$ for $|\omega^{'}_m| >|\omega_m|$. Within this approximation  we have, for $\omega_m >0$,
\begin{widetext}
\beq
\Phi (\omega_m) \approx  \frac{1-\gamma}{N} \left(\frac{1}{|\omega_m|^\gamma} \int_{0}^{\omega_m}
\frac{d\omega^{'}_m \Phi (\omega^{'}_m)}{|\omega^{'}_m|^{1-\gamma} (1+ |\omega^{'}_m|^\gamma)} + \int_{\omega_m}^{\infty}
\frac{d\omega^{'}_m \Phi (\omega^{'}_m)}{|\omega^{'}_m|^{2-\gamma} (1+ |\omega^{'}_m|^\gamma)}\right)
+ \Phi_0 (\omega_m)
 \label{eq:8}
\eeq
or
\end{widetext}
\begin{equation}\label{eq:8z}
 \Phi (z)=\left(\frac{1}{4}-b^{2} \right)\left[\frac{1}{z}\int_{0}^{z} \frac{\Phi (x)dx}{1+x}+\int_{z}^{\infty }\frac{\Phi (x)dx}{x(1+x)}  \right]
+\Phi_{0}(z)
\end{equation}
Where we introduced $z = \omega^{\gamma}_m$ and
 \beq
  b = \frac{1}{2} \sqrt{\frac{N-N_{cr}}{N}}, ~~ N_{cr} = 4(1-\gamma)/\gamma
 \label{eq:17}
  \eeq
This $N_{cr}$ coincides with \eqref{eq:10} at small $\gamma $.
Notice that for $N>N_{cr}$, $b$ is real and $b\in [0,1/2)$, while for $0<N<N_{cr}$, $b=i\tilde{b}$ is imaginary and $\tilde{b}\in [0,\infty )$. Also notice, that the equation is symmetric under $b\rightarrow -b$.
Differentiating \eqref{eq:8z} twice over $z$, we obtain second order differential equation for
 $\Phi (z)$ in the form
\beq
\frac{d}{dz} \left[z^2 \frac{d \Phi(z)}{dz}\right] + \left(\frac{1}{4}-b^{2} \right) \frac{\Phi (z)}{1+z} = \frac{d}{dz} \left[z^2 \frac{d \Phi_{0}(z)}{dz}\right]
\label{eq:9}
\eeq
We verified numerically that the solutions of the integral equation (\ref{eq:7}) and the differential equation (\ref{eq:9}) almost coincide for all $\gamma <1$.  Below we will analyze both the approximate integral equation \eqref{eq:8} (or \eqref{eq:8z})
 and the  differential equation  (\ref{eq:9}).

\section{Pairing susceptibility for $\Phi_0 (\omega_m) =
\Phi_0$}
\label{sec:3}

\subsection{BCS vs quantum-critical case}

To set the stage for our analysis, let's momentarily consider
 BCS case $\gamma =0$. In this limit, the interaction,
    the pairing vertex, and
     $\chi_{pp}$ are frequency independent. We set $V(\Omega_m) = \lambda$ and impose the upper energy cutoff for the theory at $\Lambda$.
  Because the ground state is a superconductor for any $N$, $T_c$ is non-zero and to ceck how the susceptibility evolves near the onset of pairing we need to compute $\chi_{pp}$  at a finite  $T$.
 The BCS  pairing susceptibility  is obtained in a straightforward manner by summing up Cooper logarithms and is given by
  \bea
  &&\chi_{pp} =  \nonumber \\
  &&1 +\frac{\lambda^*}{N} \log{\frac{\Lambda}{T}} + \left(\frac{\lambda^*}{N}\right)^2 \log^{2}{\frac{\Lambda}{T}} + \left(\frac{\lambda^*}{N}\right)^3  \log^{3}{\frac{\Lambda}{T}} + ... \nonumber \\
  && = \frac{1}{1-\left(\frac{\lambda^*}{N}\right) \log{\frac{\Lambda}{T}}} = \frac{N}{\lambda^*} \frac{1}{\log{\frac{T_c}{T}}}
\label{eq:1}
\eea
 where $\lambda^* = \lambda/(1+ \lambda)$ and
 $T_c = \Lambda e^{-N/\lambda^*}$.  The same holds in the Eliashberg theory for a non-critical metal, the only difference is that the upper cutoff $\Lambda$ is determined within the theory.
  The pairing susceptibility is positive at $T >T_c$,  diverges at $T=T_c$ and becomes negative at $T< T_c$, which is a well-known result.

For a quantum-critical case, $\gamma$ is non-zero.  The response to a static $\Phi_0$ is also logarithmic, because the pairing kernel is the product of the  singular interaction $1/|\omega_m - \omega^{'}_m|^\gamma$ and
$1/|\Sigma (\omega_m)| \propto 1/|\omega_m|^{1-\gamma}$. This makes the pairing kernel marginal, as in the BCS case. There is one distinction, however --   the argument of the logarithm is the running frequency $\omega_m$ rather than $T$~\cite{acf}. We can then set $T=0$ and check how the susceptibility evolves around $N_{cr}$.

  The logarithmic series for $\Phi (\omega_m)$ can be obtained by doing
    iterations, starting from $\Phi (\omega_m) = \Phi_0$.   Keeping only the highest power of the logarithm at each level of iterations, we obtain  at small $|\omega_m|$
     \bea
 && \chi_{pp} (\omega_m) =  1 + \frac{1-\gamma}{N} \log{\frac{1}{|\omega_m|}} \nonumber \\
 &&  +
   \frac{1}{2} \left(\frac{1-\gamma}{N} \log{\frac{1}{|\omega_m|}}\right)^2 + \frac{1}{6}  \left(\frac{1-\gamma}{N} \log{\frac{1}{|\omega_m|}}\right)^3 + ...
\label{eq:2}
\eea
This result holds for both the original integral equation \eqref{eq:7} and its approximate form
 \eqref{eq:8} (or \eqref{eq:8z}).
The series are similar to those in Eq. (\ref{eq:1}), but with different combinatoric factors, which
 are binomial coefficients of the Tailor expansion of the exponent. Summing up the series, we obtain
 \beq
  \chi_{pp} (\omega_m) =  e^{\frac{1-\gamma}{N} \log{\frac{1}{|\omega_m|}}} = \frac{1}{|\omega_m|^{\frac{1-\gamma}{N}}}
 \label{eq:3}
\eeq
We see that the susceptibility increases with decreasing $\omega_m$, but remains positive and non-singular for any nonzero frequency.
This is entirely expected for $N > N_{cr}$ but not for $N \leq N_{cr}$, at which the
normal state should become unstable towards pairing. Yet,  taken at a face value, Eq. (\ref{eq:3}) shows  that the pairing susceptibility remains  positive for any $N$.   This can be also cast in the renormalization group language
 (see Appendix \ref{sec:RG}).

\subsection{Quantum-critical case beyond the leading logarithms}

The apparent indifference of $\chi_{pp}$  in (\ref{eq:3})
 to $N_{cr}$
calls for extending the analysis beyond the leading logarithms.  This is what we do next.

We first notice that
the differential equation \eqref{eq:9} has the same form with and without
$\Phi_0$ (in this Section $\Phi_{0}$ is constant), hence the solutions must be the same. We can then borrow the results obtained without the source term
 and express the two linearly independent solutions of \eqref{eq:9} as
  \beq \label{eq:Phi1Phi2}
\Phi_1 (z) = H_b (z),~~ \Phi_2 (z) = H_{-b}(z)
\eeq
where
 \begin{widetext}
 \beq
 H_b (z) = \frac{1+z}{z^{1/2-b}} \frac{\Gamma\left(\frac{1}{2} +b\right)\Gamma\left(\frac{3}{2} +b\right)}{\Gamma\left(1 +2b\right)} {_2}F_{1} \left[\frac{1}{2} +b, \frac{3}{2} +b, 1 +2b, -z\right]
 \label{eq:14}
 \eeq
 \end{widetext}
  and ${_2}F{_1} [...]$ is a Hypergeometric function.  At small $z$,   $H_b (z) \sim  1/z^{1/2-b}$. At large $z$, $H_b (z) = 1+\left(\frac{1}{4}-b^{2} \right) \log{z}/z+O(1/z)$.

  A general solution of   Eq. (\ref{eq:9}) is any linear combination of $\Phi_{1}(z)$ and $\Phi_{2} (z)$.
  However, only a certain combination of $H_b (z)$ and $H_{-b} (z)$
    should generally be the solution of
     \eqref{eq:8z}, which contains $\Phi_0$ as the source term.
      To find this combination we note that because
      use
\begin{equation}
\left(\frac{1}{4}-b^{2} \right) \frac{H_{\pm b} (z)}{1+z}=- \frac{d}{dz} \left[z^2 \frac{d H_{\pm b}(z)}{dz}\right],
\end{equation}
both functions satisfy
 \begin{widetext}
\begin{eqnarray}
&&
\left(\frac{1}{4}-b^{2} \right)\frac{1}{z}\int_{0}^{z} \frac{H_{\pm} (x)dx}{1+x}
=
-\frac{1}{z}\int_{0}^{z}dx\frac{d}{dx} \left[x^2 \frac{d H_{\pm} (x)}{dx}\right]
=
-z\frac{d H_{\pm} (z)}{dz}
\\
&&
\left(\frac{1}{4}-b^{2} \right)\int_{z}^{\infty }\frac{H_{\pm} (x)dx}{x(1+x)}
=
-\int_{z}^{\infty }\frac{dx}{x}\frac{d}{dx} \left[x^2 \frac{d H_{\pm}(x)}{dx}\right]
=
z\frac{dH_{\pm}(z)}{dz}+H_{\pm} (z)- H_{\pm} (\infty)
\label{ee:16}
\end{eqnarray}
Combining and using $H_{\pm} (\infty)=1$, we find that both functions satisfy
\begin{equation}
H_{\pm} (z) = \left(\frac{1}{4}-b^{2} \right)\frac{1}{z}\int_{0}^{z} \frac{H_{\pm} (x)dx}{1+x}+ \left(\frac{1}{4}-b^{2} \right)\int_{z}^{\infty }\frac{H_{\pm} (x)dx}{x(1+x)} +1
\end{equation}
 \end{widetext}
Comparing with \eqref{eq:8z} we immediately fund that the solution of \eqref{eq:8z} is
\begin{equation}\label{eq:form}
\Phi (z) = \Phi_0 \left[H_b (z) + C(b) (H_b (z) - H_{-b} (z)) \right]
\end{equation}
where $C (b)$ is arbitrary.

 Taken at a face value, this results would imply that
 \beq
 \chi_{pp} (z) = H_b (z) + C(b) (H_b (z) - H_{-b} (z)),
 \label{eq:16}
 \eeq
i.e. any $N$, the pairing susceptibility  depends on the free parameter $C(b)$, in variance with the summation of the leading logarithms.

 We argue, however, that this is not the case, and $C (b)=0$ for  all $N> N_{cr}$.  The argument is two-fold.  First, we
 notice that at $N \to \infty$ ($ b \to 1/2$), $H_{b} (z) \to 1$, while $H_{-b} (z) \to (2+z)/z$.
 At $N = \infty$, the interaction  in the particle-particle channel vanishes, and by continuity we should get $\Phi (z) =\Phi_0$, i.e., $ \chi_{pp} (z) =1 = H_{1/2} (z)$. This implies that $C(1/2) =0$.  Second, at a finite $N$, we can obtain
 $\chi_{pp}$ in  an expansion in $\lambda = N_{cr}/N$. This expansion is reproduced by solving the integral equation \eqref{eq:8z} by iterations, i.e., expressing $\Phi (z)$ as $\Phi(z)=\Phi_{0} (1+\lambda \phi^{(1)}(z)+\lambda^{2} \phi^{(2)}(z)+\dots)$. Substituting this expansion into \eqref{eq:8z} and  performing the first two iterations, we obtain at small $z$
 \begin{eqnarray}
 &&\phi^{(1)}(z)
= -\frac{1}{4}\log z +O(1)
\nonumber\\
 &&
\phi^{(2)}(z)
= -\frac{1}{16}\log z+\frac{1}{32} \log^{2}z+O(1)
\label{eq:Phi1Eq}
\end{eqnarray}
Subsequent iterations will generate  $\log{z} $ in higher and higher powers, but also renormalize  the
  prefactors for the terms with smaller powers of  $\log{z}$.
  The full result of iterations can be expressed as
\begin{equation}\label{eq:iterationsForm}
\Phi (z)=\Phi_{0}\sum_{n=0}^{\infty } \frac{(-1)^n}{n!} a_n \log^{n}{z}.
\end{equation}
 where a straightforward algebra yields the recursive relation
\begin{equation}\label{eq:aDifference}
 a_{n+1}-a_{n}=-\frac{\lambda}{4} a_{n-1}
\end{equation}
for $n \geq 1$.
The solution consistent with $a_0 =1$, $a_1 =\lambda/4 + \lambda^2/16 + ...$, as in (\ref{eq:Phi1Eq}) is
\begin{equation}\label{eq:aGeneral}
 a_{n}=\alpha^{n},\qquad \alpha=\frac{1- \sqrt{1-\lambda }}{2} = \frac{1}{2} -b = \sum_{m=1}^\infty  D_m \lambda^m
 \end{equation}
where
\beq
D_m = \frac{(-1)^{m+1}}{2} \frac{\Gamma(3/2)}{\Gamma(3/2-m) \Gamma (m+1)}.
\eeq
Substituting into (\ref{eq:iterationsForm}), we find that at small $z$,
\begin{equation}\label{eq:PhiA}
 \Phi (z) \propto  \Phi_0/ z^{\alpha}=  \Phi_0/ z^{1/2-b}
\end{equation}
Comparing with (\ref{eq:form}) and using that at small $z$, $H_b \propto 1/z^{1/2-b}$ and $H_{-b} \propto  1/z^{1/2+b}$,
we fund that $H_{-b} (z)$ is {\it not} generated at small $z$.

 The same holds at large $z$. Here, iterations yield
 \beq
 \Phi(z) =1 + \frac{\lambda}{4} \frac{\log{z}}{z} + \frac{{\bar a}_1}{z} + ...
 \label{eq:20}
  \eeq
where
\beq
{\bar a}_1 = \sum_{m=1}^{\infty} {\bar D}_m \lambda^m
 \label{eq:21}
  \eeq
  and  ${\bar D}_1 =1/4$, $D_2 = \pi^2/48$, ...
 We explicitly verified that
 \beq
{\bar a}_1 = \frac{\lambda}{2}  \left(1 -2 \psi\left(1-\frac{1-\sqrt{1-\lambda}}{2} \right) -\psi(1)\right)
\label{eq:22}
  \eeq
where $\psi(x)$ is a di-Gamma function.
Comparing (\ref{eq:20}) and (\ref{eq:22}) with the expansion of  $H_b (z)$ and $H_{-b} (z)$ in $1/z$ we find that
(\ref{eq:20}) matches perfectly the expansion of $H_b (z)$, i.e., $H_{-b} (z)$ is not generated at large $z$ as well.
It is then natural to assume that $H_{-b} (z)$ is not generated in iterations for any $z$.   Hence, $C(b) =0$ also for $b <1/2$, and
\beq
\chi_{pp} = H_b (z)
\label{ee:12}
\eeq
 The function $H_b (z)$ is positive for all $z$, hence the pairing susceptibility is also positive.

At this stage the result for $\chi_{pp} (z)$ is consistent with the one that we obtained by summing up the leading logs -
 the only difference is that the exponent $p$ in $\chi_{pp} (z) \propto 1/z^p$ gets renormalized from  $p=\lambda$ by terms of higher-order in $\lambda$.  The susceptibility remains regular and positive even at $N = N_{cr}$ ($\lambda =1/4$, $b=0$).   Specifically, at $b=0$,
 \beq
 \chi_{pp} (z) \approx \frac{1}{z^{1/2}}
 \eeq
at  small $z$, and
 \beq
 \chi_{pp} (z) \approx 1 + \frac{1}{4} \frac{\log{z}}{z} + 0.943 \frac{1}{z} + ...
 \eeq
at large $z$. In both limits, $\chi_{pp}$ shows no indication that the system is at a critical point towards superconductivity.

\begin{figure}[!t]
\centering
\includegraphics[angle=0,width=0.5\textwidth]{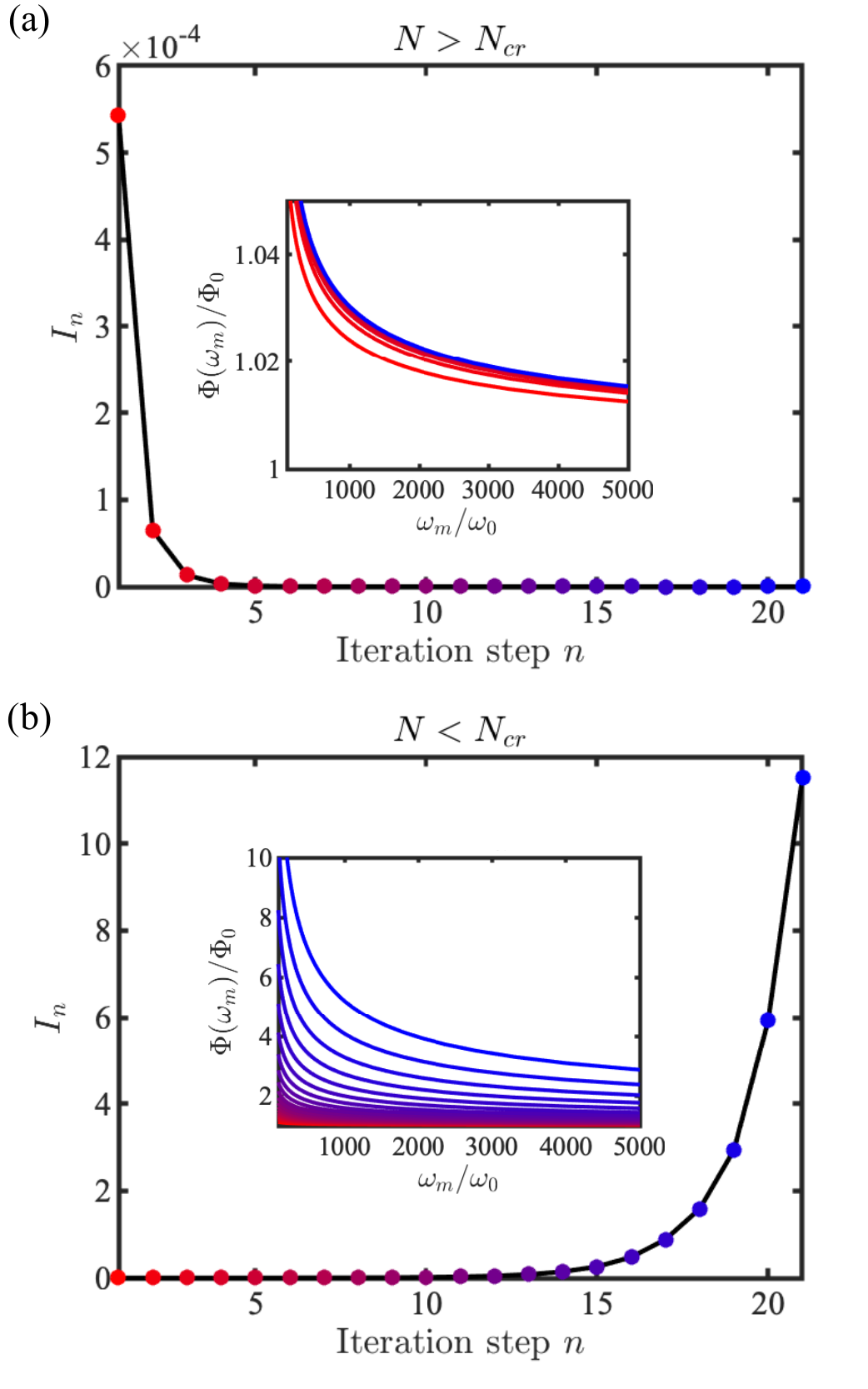}
\caption{
The iterative solution of the gap equation \eqref{eq:7} with a constant source term $\Phi(\omega_m) = \Phi_0$ and $\gamma = 1/2$. The plots show $I_{n}=\frac{1}{2\pi }\int_{0}^{\infty }\left(\Phi^{n+1}(\omega_{m})-\Phi_{n}(\omega_{m}) \right)^{2}d\omega_{m}$. For $N > N_{\text{cr}}$ (a), the iterations converge; for $N < N_{\text{cr}}$ (b), the iterations do not converge. The inserts show how the function $\Phi_{n}(\omega_{m})$ evolves with iterations.
}
\label{fig:fig2}
\end{figure}

\begin{figure}[!t]
\centering
\includegraphics[angle=0,width=0.5\textwidth]{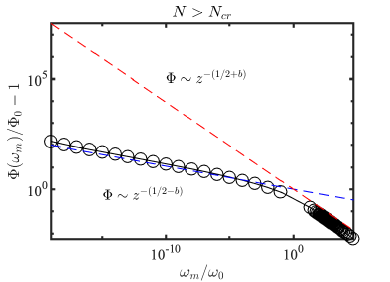}
\caption{Power-law behavior of the pairing vertex $\Phi(\omega_m)$ at small frequencies,  extracted from the iterative solution for  $\gamma=0.5$ and
$N=6.5 > N_{cr}$.
The form of $\Phi(\omega_m)$ matches the analytical result  $\Phi \sim z^{-\gamma(1/2-b)}$ from \eqref{eq:form}. }
\label{fig:fig3}
\end{figure}

\begin{figure}[!t]
\centering
\includegraphics[angle=0,width=0.5\textwidth]{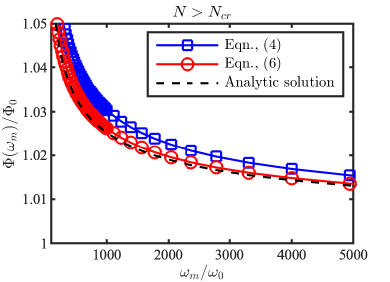}
\caption{Pairing vertex $\Phi(\omega_m)$, obtained by solving iteratively the original integral gap equation ~\eqref{eq:7} (blue line) and the approximate gap equation \eqref{eq:8} (red line) for $\gamma = 0.5$ and
$N=6.5 > N_{cr}$.
}
\label{fig:fig4}
\end{figure}

  To verify our analytical analysis, we solved the gap equation (\ref{eq:8z}) numerically using the iteration method, starting with a constant input, $\Phi(\omega_m) = \Phi_0$. We show the results in Fig.~\ref{fig:fig2}. For $N > N_{\text{cr}}$,
  the iteration procedure  converges to  a certain $\chi_{pp} (\omega_m)$ see Fig.~\ref{fig:fig2}(a). At small frequencies, this $\chi_{pp} (\omega_m)$  exhibits a clear power-law behavior  with the exponent that matches the one in  Eq.~\eqref{eq:PhiA} (see Fig.~\ref{fig:fig3}).  Across the entire frequency range,  it is quite close to the analytical expression $H_b(z)$ (see Fig.~\ref{fig:fig4}).  As an extra verification, we  solved iteratively the original integral equation (\ref{eq:7})  and also obtained a
   power-law solution at small frequencies. The
    agreement between the iterative solution of the integral gap equation and that of Eqn. (\ref{eq:8z}) is reasonably good, see  Fig.~\ref{fig:fig4}.

What happens at smaller $N$, i.e., larger $\lambda = N_{cr}/(4N)$?  To answer, we note that our results for $\Phi (z)$ and $\chi_{pp} (z)$ have been obtained by iterations. This procedure is valid as long as the  iterations converge. Let's  check the convergence of the series in Eqs. (\ref{eq:aGeneral}) and (\ref{eq:21}).
  For this, we need the forms of $D_n$ and ${\bar D}_n$ at large $n$.
 Using the expressions for  Gamma  and di-Gamma functions at large argument, we find  that $D_n$ and ${\bar D}_n$  decrease at large $n$,  but {\it by a power-law rather than exponentially}. In explicit form,
 \beq
 D_n \approx \frac{0.141}{n^{3/2}},~~{\bar D}_n \approx \frac{0.347}{n^{3/2}}
  \label{eq:23}
  \eeq
The $1/n^{3/2}$ dependence ensures that the iterations converge for $\lambda \leq 1$, i.e., for $N \geq N_{cr}$.
However, once $N$ gets smaller than $N_{cr}$, the series diverge for both small and large $z$ because
$\lambda^n = e^{n \log{\lambda}}$ overshoots $1/n^{3/2}$ decay of $D_n$ and ${\bar D}_n$.
 In Fig.~\ref{fig:fig2}(b)
 we show our numerical results, which confirm that for $\lambda >1$, iteration series  diverge for all $z$

To obtain the actual $\chi_{pp} (z)$ for $N<N_{cr}$, we note that its generic form is
 still a linear combination of $H_b (z)$ and $H_{-b} (z)$, but now
 the argument $b = (1/2) \sqrt{1-\lambda}$ becomes imaginary: $b \to i\tilde{b}$, where
\beq
\tilde{b} = \frac{1}{2} \sqrt{\lambda -1}
\eeq
In this situation, $H_b (z)$ and $H_{-b} (z)$ become complex conjugated functions $H_{i\tilde{b}} (z)$ and $H_{-i\tilde{b}} (z)$, and {\it {both must be kept in
(\ref{eq:16}) to ensure that $\Phi(z)$ is a real function of frequency}}.  A generic real solution of Eq.
(\ref{eq:8})
 is
\beq
\Phi (z) = \Phi  (z, \eta) = \Phi_0  \frac{H_{i\tilde{b}} (z) e^{i\eta} + H_{-i\tilde{b}} (z) e^{-i\eta}}{2 \cos{\eta}}
\label{eq:24a}
\eeq
where $\eta$ is arbitrary within $0 < \eta <\pi$. There is no condition on $\eta$ as Eq. (\ref{eq:24}) does not have to match with perturbation theory.  Accordingly,
\beq
\chi_{pp}  (z) = \chi_{pp} (z, \eta) = \frac{H_{i\tilde{b}} (z) e^{i\eta} + H_{-i\tilde{b}} (z) e^{-i\eta}}{2 \cos{\eta}}
\label{eq:24}
\eeq
We see that the pairing susceptibility becomes the  function not only of $z$ but also of a free parameter $\eta$.
 To see the outcome, set ${\tilde b}$ to be small (i.e., set $N$ to be slightly below $N_{cr}$) and consider large $z$. Using the forms of $H_{\pm i\tilde{b}} (z)$ at small index and large argument, we obtain
\beq
\chi_{pp} (z, \eta) = 1 + \frac{1}{4} \frac{\log{z}}{z} + \frac{0.943}{z} \left(1 + c \tilde{b} \tan{\eta}\right)
\label{eq:25}
\eeq
 where  $c= \pi^2/(4*0.943) =2.61$
We see that for any non-zero $\tilde{b}$, there is a range of $\eta \geq \pi/2$, where
\beq
\pi^2 {\bar b} |\tan{\eta}| > \log{z}
\eeq
 and the pairing susceptibility is negative.  This is a clear indication that the system is now below a pairing instability.  At the same time, if we choose $\eta = \pi/2$, we find that the pairing susceptibility is infinite. The latter holds for any value of $\tilde{b}$, i.e. for  any $N < N_{cr}$.  An infinite pairing susceptibility normally  implies that the system is right at the onset of a superconducting order.
 This is very different from BCS case where the pairing susceptibility is a single-valued function, singular at $T_c$ and negative at $T <T_c$.

 This highly unconventional behavior of $\chi_{pp} (z, \eta)$ is fully consistent with the
  existence of an infinite set of solutions of the non-linear gap equation at $N < N_{cr}$. As we stated in the Introduction, earlier studies have found that in the absence of $\Phi_0$, there exists an infinite set of solutions for the pairing vertex
  $\Phi_n (z)$. The solutions are parametrized by $n$ and the overall magnitude of $\Phi_n (\omega_m)$ decreases with increasing $n$. In the limit $n =\infty$, $\Phi_n (z)$ is exponentially small and is the solution of the linearized gap equation. In explicit form,
 $\Phi_\infty (z) = i P (H_{i\tilde{b}} (z) -  H_{-i\tilde{b}} (z))$, where
  $P$ is an infinitesimal real number.
    \footnote{A more accurate statement is $\Phi_\infty (z)$ is the limiting form of the solution of the non-linear gap equation at a vanishingly small but still finite $\Phi(z)$. The distinction with the solution of just the linearized gap equation is an infinitesimally small $\Phi_{\infty}(z)$ is essential because  $\Phi_{\infty} (z)$ is formally non-normalized: $\Phi^2_\infty (z)$  behaves as $1/z$ at small $z$, and the corresponding condensation energy diverges logarithmically at $z \to 0$. A finite $\Phi (z)$ tends to finite value at $z =0$ and regularizes the singularity. Viewed as the limit of $\Phi (z) \to 0$, $\Phi_{\infty}(z)$ then belongs to the class of normalized functions.}. 
      Comparing this with our result that the pairing susceptibility depends on a free  parameter $\eta$,  we see
   that the existence of a range of $\eta$ where
  $\chi_{pp} (z, \eta)$ is negative, is consistent with the presence of  infinite set of different ordered states, while the fact that  $\chi_{pp} (z, \pi/2)$ is infinite  is consistent with the still existence of the solution of the linearized gap equation for all  $N < N_{cr}$.   
  
  We note in passing that there is a certain similarity between the solution of the linearized gap equation at $N < N_{cr}$ in our case and the
   solution of the relativistic Klein-Gordon/Dirac equation for a heavy atom with an atomic number
    $Z$ larger than a half of the inverse fine structure constant for a state with the angular momentum $l=0$ (see, e.g.,
    \cite{Zeld_Popov,Schwabl} and \cite{Zohar} for a more recent analysis).
   In both cases the solution (the gap function $\Phi (z)$ in our case and the wave function $\Phi (r)$ for a heavy atom) oscillates at small argument as a function of the logarithm of the argument.  Also in both cases a solution
    is 'weakly" non-normalizable, as it gives rise to logarithmically divergent condensation energy in our case and kinetic energy of an atom.  The  regularization is provided by non-linearity in our case and 
     a finite atomic size the case of 
    a heavy atom.
      This analogy clearly deserves  further study.

\section{Pairing susceptibility for  frequency-dependent $\Phi_0 (\omega_m)$}
\label{sec:4}

For a constant $\Phi_0$ as the source term, we found that the pairing susceptibility displays a highly unconventional behavior: it does diverge as $N$ approaches $N_{cr}$ and remains finite and positive at $N=N_{cr}$,
   but at smaller $N$ becomes a function of a parameter $\eta$ and diverges at $\eta = \pi/2$.
    We now analyze whether this behavior is specific to a constant $\Phi_0$ or persists for frequency-dependent
     $\Phi_0 (\omega_m)$.  As before, we will use $z= \omega_m^\gamma$ instead of $\omega_m$.
We first consider a particular set of $\Phi_0(z)$ and then discuss an arbitrary $\Phi_0 (z)$. For a set, consider
\beq
\Phi (z) = \Phi_0 (1 + \epsilon H_{b^*} (z))
\label{ee_17}
\eeq
 where $b^*$ is positive and real, but different from $b =(1/2) \sqrt{1-N_{cr}/N}$.

 We will be searching for the solution of the gap equation (\ref{eq:8})  in the form
 \beq
 \Phi (z) = \Phi_0 \left(c_1 H_b (z) + c_2 H_{b^*} (z)\right)
 \eeq
  substituting this form into the r.h.s of (\ref{eq:8}) and using (\ref{ee:16}), we obtain after a simple algebra the set of two equations
  \beq
  c_1 =1-dc_2, ~~c_2 (1 -d) =\epsilon,
  \eeq
  where
  \beq
  d = \frac{\frac{1}{4} -b^2}{\frac{1}{4} -(b^*)^2}
  \eeq
  The susceptibility
  \beq
  \chi_{pp} (z) = (1+ \epsilon) \left(H_b + \frac{\epsilon}{1+ \epsilon} \frac{H_{b^*} (z) - H_b (z)}{1-d}\right)
  \eeq
  For a generic $b^* \neq b$, the susceptibility displays the same behavior as for a constant $\Phi_0$, i.e., it remains regular and positive for $N \geq N_{cr}$ and becomes a  function of a free parameter $\eta$  at $N < N_{cr}$.
  The new behavior emerges for a class of $\Phi_0 (z)$ from (\ref{ee_17}) with $b^* = \alpha b$ and $\alpha$ positive. In this case, $1-d = b^2 (1-\alpha^2) /(1/4 -\alpha^2 b^2)$ and at $H_{b^*} (z) - H_b (z) = -b (1-\alpha) dH_b (z)/db + O(b^2)$. Expressing  $2b = \sqrt{1-\lambda}$, we then obtain at small $b$
   \beq
  \chi_{pp} (z) = (1+ \epsilon) \left(H_b -\frac{1}{2(1+ \alpha)} \frac{\left. dH_b (z)/db\right|_{b=0}}{\sqrt{1-\lambda}}\right)
 \label{ee_19}
  \eeq
 We see that for these $\Phi_0 (z)$,  the pairing susceptibility does diverge as $1/\sqrt{N-N_{cr}}$ when
  $N$ approaches $N_{cr}$ and $\lambda = N_{cr}/N$ approaches $1$.
   In explicit form
   $\left.\frac{\partial H_{b}(z)}{\partial b}\right|_{b=0}=G_{2,2}^{2,0}\left(-z\left|
\begin{array}{cc}
 0,&1 \\
 -1/2,&-1/2 \\
\end{array}
\right.\right)$, where $G^{mn}_{pq}$ is the Meijer $G$ function. It behaves as $\sim\frac{\log z}{\sqrt{z}}$ at  $z\rightarrow 0$ and as $\sim \frac{1}{z}$ at $z\rightarrow \infty $.

 This result is not surprising as
  at $N = N_{cr} -0$,  the solution of the linearized gap equation without the source term  is
  \beq
  \Phi (z) = a_{1} H_{0}(z) +  b_{2} \left.dH_b (z)/db\right|_{b=0}
  \label{ee_18}
  \eeq
   with arbitrary $a_{1}$ and $a_{2}$.  From this perspective, the appearance of
   $\left.dH_b (z)/db\right|_{b=0}$ in (\ref{ee_19}) with the diverging prefactor at $N \to N_{cr}$, is
     a clear indication that the system is about to develop spontaneously a non-zero $\Phi (z)$ at $N < N_{cr}$.

These results can be extended to a generic  $\Phi_0 (z)$. We show in the next Section that
$\Psi_{\tilde b} = iA (H_{i{\tilde b}} (z) -H_{-i{\tilde b}} (z))$ form a complete set of orthogonal functions which, with a proper normalization factor $A$,  satisfy
 \begin{equation}\label{eq:normalization}
\langle \Psi_{\tilde{b}}| \Psi_{\tilde{b}'}\rangle=2\pi \delta (\tilde{b}-\tilde{b'}),
\end{equation}
An arbitrary $\Phi_0 (z)$ can then be expressed as
\beq
\Phi_0 (z) = \int_0^{\infty} d {\tilde b} C_{\tilde b} \Phi_{\tilde b} (z)
\label{ee_20}
  \eeq
Whether the pairing susceptibility remains finite or diverges at $N =N_{cr}$ depends on $C_{0}$. If it vanishes, the pairing susceptibility remains finite at $N =N_{cr}$, while it is non-zero, the susceptibility  diverges at $N \to N_{cr}$ as $1/\sqrt{N-N_{cr}}$.

\section{Physical reasoning}
\label{sec:5}

We now go back to the issue about the number of solutions of the gap equation with the external source $\Phi_0 (z)$. We recall that Eq. \eqref{eq:8z} has two linearly independent solutions for any $N$, see (\ref{eq:form}), yet only one solution is reproduced by  iterations at $N>N_{cr}$.  A natural question then is whether there a
  physical reasoning for choosing the one-component solution selected by the iterations and not a general two-component solution.   In this section we give such a  reasoning using the analogy between our gap equation and Schrodinger equation in quantum mechanics.

To do this, we go back to the basics of Eliashberg theory and recall the requirement that the condensation energy $E[\Phi (\omega_{m})]$ for a given $\Phi (\omega_m)$  must be finite, otherwise the probability to find the state with this $\Phi (\omega_m)$ vanishes.
  For infinitesimally small $\Phi (\omega_m)$, which we consider in this paper,
  the condensation energy $E_c$ is quadratic in $\Phi (\omega_{m})$  and is given by
  \beq
  E_c = A \int_0^{\infty} \frac{\Phi^2 (\omega_{m}) d\omega_{m} }{|\omega_{m} |^{1-\gamma }(1+|\omega_{m}|^{\gamma } )}.
  \eeq
 where $A$ is a finite overall factor.

 In analogy with quantum mechanics,  the  gap equation can be re-expressed as
    the effective
    Schroedinger-type equation $\Phi (\omega_m) = \hat{L}\Phi(\omega_{m}) + \Phi_0 (\omega_m)$ with the linear operator
   \begin{equation}\label{eq:L}
\hat{L}\Phi(\omega_{m}) =\int \frac{d\omega^{'}_m \Phi (\omega^{'}_m)}{|\omega^{'}_m|^{1-\gamma} |\omega_m -\omega^{'}_m|^\gamma} \frac{1}{1+ |\omega^{'}_m|^\gamma}
\end{equation}
Then $\Phi (\omega_{m})$ can be viewed as a
   (real) wavefunction,  and the condensation energy,
    or, equivalently,  the  scalar  product in the space of wavefunctions  $\langle \Phi | \Psi \rangle$
     can be viewed as the norm. The scalar product is
    defined as
 \begin{equation}\label{eq:dotOmega}
 \langle \Phi | \Psi \rangle=\int_0^{\infty} \frac{\Phi (\omega_{m})\Psi (\omega_{m} )d\omega_{m} }{|\omega_{m} |^{1-\gamma }(1+|\omega_{m}|^{\gamma } )}.
\end{equation}
and the
 condensation energy is
\beq
E_c = A  \langle \Phi | \Phi \rangle
\label{ee_21}
\eeq
 One can immediately check that
for the scalar product defined in (\ref{eq:dotOmega})
 $\langle \Phi |\hat{L}\Psi \rangle = \langle \hat{L}\Phi |\Psi \rangle $, i,e., the linear operator ${\hat L}$ is Hermitian, as the physical operator should be.  Physically meaningful $\Phi (\omega_m)$  then should have a  finite norm.
For the truncated integral equation  \eqref{eq:8z} we introduce $z = \omega^\gamma_m$ and
\begin{equation}\label{eq:Pi}
\hat{\Pi }\Phi(z) = \frac{1}{z}\int_{0}^{z} \frac{\Phi (x)dx}{1+x}+\int_{z}^{\infty }\frac{\Phi (x)dx}{x(1+x)}
\end{equation}
 and re-express \eqref{eq:8z} as
 \begin{equation}\label{eq:8zPi}
 \Phi(z) =\left(\frac{1}{4}-b^{2} \right)\hat{\Pi }\Phi(z) +\Phi_{0}(z)
\end{equation}
 One can verify that $\hat{\Pi }$ is also  Hermitian for the scalar product defined in
\eqref{eq:dotOmega}, so   physically meaningful $\Phi (z)$
 should have a finite norm.

For  $\Phi_{0}(z)=\Phi_{0}$, there are two eigenfunctions
$\Phi_{1} (z) = H_b (z)$ and $\Phi_{2}(z)= H_{-b} (z)$, see  \eqref{eq:Phi1Phi2}.
Substituting both into (\ref{ee_21}) and analyzing the
  behavior at $z\rightarrow 0$,
   we see that $H_{b}(z)$ has a finite norm, while the norm of $H_{-b}(z)$ diverges. Hence, only $H_b (z)$
    is physically meaningful. This is physical justification for choosing $\Phi (z) = \Phi_0 H_b (z)$ based on the iteration procedure.
\footnote{The norm of both functions formally  diverges logarithmically at $z\rightarrow \infty $. This divergence is artificial as it  appears because the norm of $\Phi_{0}$ diverges, i.e. formally
 $\Phi_0$ is outside the space of allowed wavefunctions. To  fix this one should instead consider $\Phi_{0}(z)=\Phi_{0}e^{-\delta z}$, where $\delta \rightarrow +0$ is the convergence factor.}

We can analyze the equation \eqref{eq:8zPi} further. Let's set $\Phi_{0}(z)=0$. The gap equation becomes
\begin{equation}\label{eq:eigen}
\hat{\Pi }\Phi_{\epsilon } =\epsilon_{b} \Phi_{\epsilon }
\end{equation}
where $\epsilon_{b}=\left(1/4-b^{2} \right)^{-1}$. This is an equation for eigenvalues/eigenfunctions of the operator $\hat{\Pi }$ in the space of wavefunctions with the norm defined in \eqref{eq:dotOmega}. As $\hat{\Pi }$ is Hermitian, all eigenvalues are real.
The solution of (\ref{eq:eigen}) is formally
 $\Phi(z) \propto \left(H_{b}(z)-H_{-b}(z) \right)$ for any $b$, but the normalizable solution appears only at $N < N_{cr}$, when
 $b=i\tilde{b}$ is imaginary.  Then the eigenvalue is $\epsilon_{b}=\left(1/4+\tilde{b}^{2}\right)^{-1}$ and
 the physically meaningful eigenfunction is
\begin{equation}\label{eq:eigenFV}
 \Phi_{\tilde{b}}=iA\left(H_{i\tilde{b}}(z)-H_{-i\tilde{b}}(z) \right),
\end{equation}
where
$A$ is the normalization factor.

We see that the operator $\hat{\Pi }$ has a continuous spectrum $\epsilon \in [0,1/4]$. The eigenfunctions must then be normalized by
\begin{equation}\label{eq:normalization_1}
\langle \Phi_{\tilde{b}}| \Phi_{\tilde{b}'}\rangle=2\pi \delta (\tilde{b}-\tilde{b'})
\end{equation}

We can then introduce the Green's function $G(z,z')$ as
\begin{equation}\label{eq:Greens}
G_b(z,z')=\int_{0}^{\infty } \frac{\Phi_{\tilde{b}}(z)\Phi_{\tilde{b}} (z')}{\left(\frac{1}{4}-b^{2} \right)^{-1}-\epsilon_{\tilde{b}}}\frac{d\tilde{b}}{2\pi } .
\end{equation}
and write the solution of the gap equation \eqref{eq:8z} with the source term $\Phi_0 (z)$ as
\begin{equation}\label{eq:G}
\Phi (z)=\left(\frac{1}{4}-b^{2} \right)^{-1}\int_{0}^{\infty } G_b(z,z')\Phi_{0}(z')\frac{dz'}{1+z'}
\end{equation}
For $b >0$ ($N > N_{cr}$) , $G_b(z,z')$ is regular and $\Phi (z)$ is single-valued. The susceptibility
$\chi_{pp} (z)$ is a regular (non-divergent) function of $z$.

For $b=i\tilde{b}$ ($N<N_{cr}$),
 the integral in \eqref{eq:Greens}
 diverges.
 The solution then becomes
\begin{equation}\label{eq:}
\Phi (z)=\Phi_{p}(z)+{\bar \eta}\Phi_{\tilde{b}}
\end{equation}
where $\Phi_{p}(z)$ is a  particular normalizable solution of \eqref{eq:8z}
and ${\bar \eta}$ is an arbitrary (real) number. The pairing susceptibility then becomes a function
 of a free running parameter ${\bar \eta}$, in agreement with the analysis in the previous Section.


\begin{figure*}[!t]
\centering
\includegraphics[angle=0,width=0.9\textwidth]{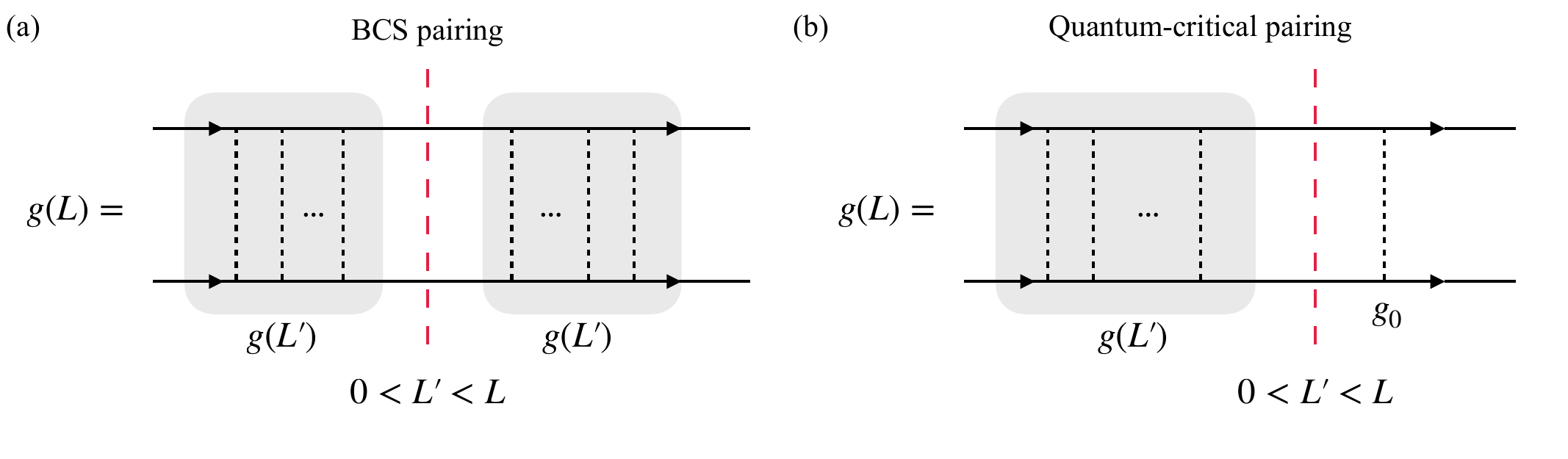}
\caption{An illustration of the structure of the  one-loop RG equation for the 4-fermion interaction vertex for
  (a) BCS pairing in a Fermi liquid and (b) quantum-critical pairing out of a non-Fermi liquid. }
\label{fig:fig5}
\end{figure*}

\section{Conclusions}
\label{sec:6}
In this paper we  analyzed the dynamical pairing susceptibility $\chi_{pp} (\omega_m)$ (the ratio of the fully dressed dynamical pairing vertex $\Phi (\omega_m)$ and the bare $\Phi_0 (\omega_m)$) at zero temperature in a quantum-critical metal, where superconductivity
  emerges out of a non-Fermi liquid ground state once the pairing interaction exceeds a certain threshold.   We showed that this susceptibility is qualitatively different from that for superconductivity emerging out of a Fermi liquid. There, the pairing susceptibility is positive above the transition, diverges at the transition, and becomes negative below it. In the quantum-critical case, we found for a static $\Phi_0$ that $\chi_{pp} (\omega_m)$ remains positive and non-singular all the way up to a pairing instability, and becomes
  a function of both  $\omega_m$ and a free parameter $\eta$ immediately below the instability.
   We argued that this highly unconventional  behavior of $\chi_{pp} (\omega_m)$  reflects a multi-critical nature of a $T=0$ onset point of  superconductivity  in a quantum-critical metal when  an infinite number of superconducting states emerges
   simultaneously with different amplitudes of the order parameter, down to an infinitesimally small one.
     We discussed how the pairing susceptibility behaves for a generic dynamical $\Phi_0 (\omega_m)$ and established the conditions when $\chi_{pp} (\omega_m)$ remains finite at the critical point and when it diverges at criticality.  We also presented physical reasoning based on the analogy between the gap equation at the critical point and
     Schrodinger-type equation in quantum mechanics.

  \begin{acknowledgments}
We acknowledge with thanks useful conversations with L. Classen, H. Goldman, R. Fernandes, E. Fradkin, M. Pospelov, J. Schmalian, D. Son, A. Vainstein, Y. Wang, and Y. Wu.
 The work by A.V.C. is supported by the NSF-DMR Grant No.2325357.
 \end{acknowledgments}

\appendix

\section{Pairing susceptibility in the renormalization group treatment}
\label{sec:RG}

In this Appendix we show how the pairing susceptibility can be obtained within the renormalization group (RG) formalism.  For this, one has to analyze ladder renormalizations of the 4-fermion pairing interaction.

In BCS theory, the bare value of the pairing interaction is
$g_0 =\lambda^*/N$, and the running one, $g (L)$, is a function of $L = \log{\Lambda/T}$. The  one-loop RG equation is obtained by (i) selecting a cross-section in the ladder series, in which  intermediate $L'$ are set to be larger than in other cross-sections, (ii)  expressing the momentum/frequency integral in this cross-section as $\int_0^L d L'$,  and (iii) summing up the renormalizations of $g$ on both sides of this cross-section
 over intermediate $L''$ up to $L'$ (Fig.~\ref{fig:fig5}(a)).
Each of such renormalizations yields $g(L')$, and the full ladder renormalization is expressed as
\beq
g(L) = \int_0^L dL' g^2 (L'), ~~{\text i.e}~~ \frac{d g(L)}{dL} = g^2 (L)
\eeq
Solving this equation one obtains for the susceptibility $\chi_{pp} (L) = g(L)/g_0$
\beq
\chi_{pp}(L) = \frac{1}{1- g_0 L} = \frac{1}{1-\left(\frac{\lambda^*}{N}\right) \log{\frac{\Lambda}{T}}}
\label{eq:1a}
\eeq
 which is the same as $\chi_{pp}$ in (\ref{eq:1}).

For quantum-critical pairing,  the RG analysis is applied most naturally to 4-fermion vertex with small incoming frequencies $(\omega_m , -\omega_m)$  and much larger outgoing frequencies of order one (all frequencies in unites of $\omega_0$).  The bare coupling is $g_0 = (1-\gamma)/N$ and the running $g(L)$ is a function of $L = \log(1/|\omega_m|)$. The one-loop RG equation is again obtained by choosing the cross-section with the largest $L'$. The momentum/frequency integral in this cross-section  yields $\int_0^L d L'$, as in the BCS case.  The difference with BCS comes about because this cross-section is now at the boundary rather than in the middle
(Fig.~\ref{fig:fig5}(b)),  as one can explicitly verify~\cite{acf}.  As a result, the renormalizations in other cross-sections yield $g_0 g(L)$ rather than $g^2 (L)$.
We then obtain
\beq
g(L) = g_0 \int_0^L dL' g (L'), ~~{\text i.e}~~ \frac{d g(L)}{dL} = g_0 g (L)
\eeq
Solving this equation, one obtains the susceptibility $\chi_{pp} (L) = g(L)/g_0$
\beq
\chi_{pp} (L) = e^{g_0 L} =  \frac{1}{|\omega_m|^{\frac{1-\gamma}{N}}},
\label{eq:2a}
\eeq
 which is the same as (\ref{eq:3}).

\bibliography{paper_chi}
\end{document}